\documentclass[twocolumn,superscriptaddress,letter]{revtex4}%
\usepackage{amssymb}
\usepackage{color}
\usepackage{graphicx}
\usepackage{dcolumn}
\usepackage{bm}
\usepackage[header,title,page,titletoc]{appendix}
\usepackage{amsmath}
\usepackage{amsfonts}%
\providecommand{\U}[1]{\protect \rule{.1in}{.1in}}

\begin{document}
\title{Topological Superconductors in Correlated Chern Insulators}
\author{Ying Liang}
\affiliation{Department of Physics, Beijing Normal University, Beijing, 100875, P. R. China}
\author{Jing He}
\affiliation{Department of Physics, Beijing Normal University, Beijing, 100875, P. R. China}
\author{Ya-Jie Wu}
\affiliation{Department of Physics, Beijing Normal University, Beijing, 100875, P. R. China}
\author{Ying-Xue Zhu}
\affiliation{Department of Physics, Beijing Normal University, Beijing, 100875, P. R. China}
\author{Su-Peng Kou}
\thanks{Corresponding author}
\email{spkou@bnu.edu.cn}
\affiliation{Department of Physics, Beijing Normal University, Beijing, 100875, P. R. China}

\begin{abstract}
In this paper, we realize a topological superconductor (TSC) in correlated
topological insulator - the interacting spinful Haldane model. We consider the
electrons on the Haldane model with on-site negative-U interaction and then
study its properties by mean field theory and random-phase-approximation (RPA)
approach. We found that in the intermediate interaction region, the ground
state becomes a TSC with the Chern number $\pm2$. We also study its edge
states and the zero modes of the $\pi$-flux.

\end{abstract}
\maketitle

\section{Introduction}

Different from the traditional superconductor (SC), topological superconductor
(TSC) always has the topologically-protected gapless Majorana edge
states\cite{vol}. For TSCs, people cannot use the local order parameter to
characterize them. Instead, it is the topological invariable that plays the
role to classify the topological properties of TSCs. According to the
characterization of "ten-fold way" from random matrix\cite{ry,kia,ryu}, there
exist three types of TSCs in two dimensions: D-type chiral TSC, of which the
topological invariable is the Chern number\cite{read}, C-type chiral TSC, of
which the topological invariable is also the Chern number and DIII-type
TSC\cite{read}, of which the topological invariable is $%
\mathbb{Z}
_{2}$ topological invariable. For the D-type chiral TSC, a typical example is
two dimensional $p_{x}\pm \mathrm{i}p_{y}$ chiral $p$-wave superconductor. This
TSC has exotic topological properties, such as the topological protected
chiral Majorana edge states and Majorana zero mode on $\pi$-flux\cite{vol}. In
particular, a $\pi$-flux with Majorana mode becomes non-Abelian anyons due to
its non-Abelian statistics\cite{read}. To realize a D-type TSC people always
consider the system with s-wave pairing SC and strong spin-orbital coupling in
strong Zeeman field\cite{so,chu}. For the C-type chiral TSC, a typical example
is two dimensional $d_{x}\pm \mathrm{i}d_{y}$ chiral $d$-wave superconductor.
This TSC also has the topological protected chiral edge states and zero mode
on $\pi$-flux\cite{read}. However, till now people cannot realize the
$d_{x}\pm \mathrm{i}d_{y}$ chiral $d$-wave superconductor in a physical system.

In this paper we will show another road to realize a TSC in an interacting
topological insulator - the interacting spinful Haldane model. The Haldane
model is a lattice model that illustrates integer quantum Hall effect without
Landau levels\cite{Haldane}. We consider the electrons on the Haldane's model
with on-site negative-U interaction and then study its properties by mean
field theory and RPA\ approach. We found that in the intermediate interaction
region, the ground state becomes a TSC with the Chern number $\pm2$. For the
ground state there is a local order parameter and the elementary excitations
are gapless phase fluctuations and gapped quasi-particle (an electron or a
hole). To characterize its topological properties, we study its edge states
and the zero modes on the $\pi$-flux. Thus we propose that the s-wave
topological superconductor/superfluid may be realized by putting two-component
(two pseudo-spins) interacting fermions on a honeycomb optical lattice.

The paper is organized as below. In Sec. II, we start with the Hamiltonian of
the interacting spinful Haldane model on the honeycomb lattice. In Sec. III,
We calculate the superconducting order parameter with mean field approach and
get a global phase diagram at zero temperature. In Sec. IV, we discuss the
topological properties of the TSC, including the zero modes of the $\pi$-flux
and the edge states. In Sec. V, we calculate the phase stiffness of the TSC.
In Sec. VI, we discuss the physical realization of the s-wave topological
superconductor/superfluid in a honeycomb optical lattice. Finally, the
conclusions are given in Sec. VII.

\section{The Spinfull Haldane model with attractive interaction}

In this paper, we consider the Hamiltonian of the correlated Chern insulator -
spinful Haldane model on honeycomb lattice with the on-site attractive
interaction as\cite{Haldane,he1,he2,he3}
\begin{equation}
H=H_{\mathrm{H}}+H^{\prime}-U\sum \limits_{i}\hat{n}_{i\uparrow}\hat
{n}_{i\downarrow}-\mu \sum \limits_{\left \langle i,\sigma \right \rangle }\hat
{c}_{i\sigma}^{\dagger}\hat{c}_{i\sigma}.
\end{equation}
where $U>0$ is the on-site interaction strength. $\hat{n}_{i\uparrow}$ and
$\hat{n}_{i\downarrow}$ are the number operators of electrons with up-spin and
that with down-spin, respectively. $\sigma$ are the spin-indices representing
spin-up $(\sigma=\uparrow)$ and spin-down $(\sigma=\downarrow)$ for electrons.
$\mu$ is the chemical potential and we only consider half-filling case by
setting $\mu=0$. $H_{\mathrm{H}}$ is the Hamiltonian of the spinful Haldane
model which is given by%
\[
H_{\mathrm{H}}=-t\sum \limits_{\left \langle {i,j}\right \rangle ,\sigma}\left(
\hat{c}_{i\sigma}^{\dagger}\hat{c}_{j\sigma}+h.c.\right)  -t^{\prime}%
\sum \limits_{\left \langle \left \langle {i,j}\right \rangle \right \rangle
,\sigma}e^{i\phi_{ij}}\hat{c}_{i\sigma}^{\dagger}\hat{c}_{j\sigma}+h.c.
\]
Here $t$ and $t^{\prime}$ are the nearest neighbor and the next nearest
neighbor, respectively. To break time-reversal symmetry, we introduce a
complex phase $\phi_{ij}$ into the second-neighbor hopping, and set the
direction of the positive phase is clockwise $\left(  \left \vert \phi
_{ij}\right \vert =\frac{\pi}{2}\right)  $. $H^{\prime}$ denotes an on-site
staggered energy which is
\[
H^{\prime}=\varepsilon \sum \limits_{i\in{A,}\sigma}\hat{c}_{i\sigma}^{\dagger
}\hat{c}_{i\sigma}-\varepsilon \sum \limits_{i\in{B,}\sigma}\hat{c}_{i\sigma
}^{\dagger}\hat{c}_{i\sigma}.
\]
\ 

\begin{figure}[ptbh]
\includegraphics[width = 8.0cm]{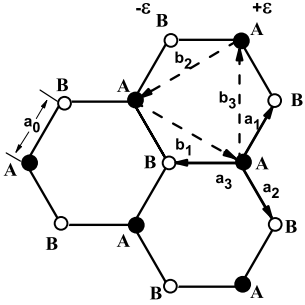}\caption{(Color online) The
illustration of the honeycomb lattice}%
\label{hon}%
\end{figure}

Using the Fourier transformations, the electronic annihilation operators on
the two sublattices are written into
\begin{align}
\hat{c}_{i\in A,\sigma}  &  =\frac{1}{\sqrt{N_{s}}}\sum_{\mathbf{k}%
}e^{i\mathbf{k}\cdot \mathbf{R}_{i}}\hat{a}_{\mathbf{k\sigma}},\nonumber \\
\hat{c}_{i\in B,\sigma}  &  =\frac{1}{\sqrt{N_{s}}}\sum_{\mathbf{k}%
}e^{i\mathbf{k}\cdot \mathbf{R}_{i}}\hat{b}_{\mathbf{k\sigma}}.
\end{align}
$N_{s}$ denotes the number of unit cells. For free fermions (the on-site
Coulomb repulsion $U$ is zero), the spectrum is
\begin{equation}
\mathbf{E}_{\mathbf{k}}=\pm \sqrt{\left \vert \xi_{k}\right \vert ^{2}+\left(
\gamma_{k}+\varepsilon \right)  ^{2}} \label{4}%
\end{equation}
where
\begin{align}
\left \vert \xi_{\mathbf{k}}\right \vert  &  =\left \vert t\sum_{i=1}%
^{3}e^{i{\mathbf{k}\cdot}\mathbf{a}_{i}}\right \vert \\
&  =t\sqrt{3+2\cos{(\sqrt{3}k_{y})}+4\cos{(3k_{x}/2)}\cos{(\sqrt{3}k_{y}/2)}%
}\nonumber
\end{align}
and
\[
\gamma_{k}=2t^{\prime}\sum \limits_{i}\sin{(\mathbf{k}\cdot \mathbf{b}_{i})}.
\]
The parameters $\mathbf{a_{1},a_{2}}$ and $\mathbf{a_{3}}$ are the nearest
neighbors of the A sublattice (See Fig.1) which are defined as
\begin{equation}
\mathbf{a}_{1}=a_{0}(\frac{1}{2},\frac{\sqrt{3}}{2}),\mathbf{a}_{2}%
=a_{0}(\frac{1}{2},-\frac{\sqrt{3}}{2}),\mathbf{a}_{3}=a_{0}(-1,0)
\end{equation}
and $\mathbf{b}_{1}=\mathbf{a}_{2}-\mathbf{a}_{3},\mathbf{b}_{2}%
=\mathbf{a}_{3}-\mathbf{a}_{1},\mathbf{b}_{3}=\mathbf{a}_{1}-\mathbf{a}_{2}.$
$a_{0}$ is the length of the hexagon side and is set to be unit.

\begin{figure}[ptbh]
\includegraphics[width = 8.0cm]{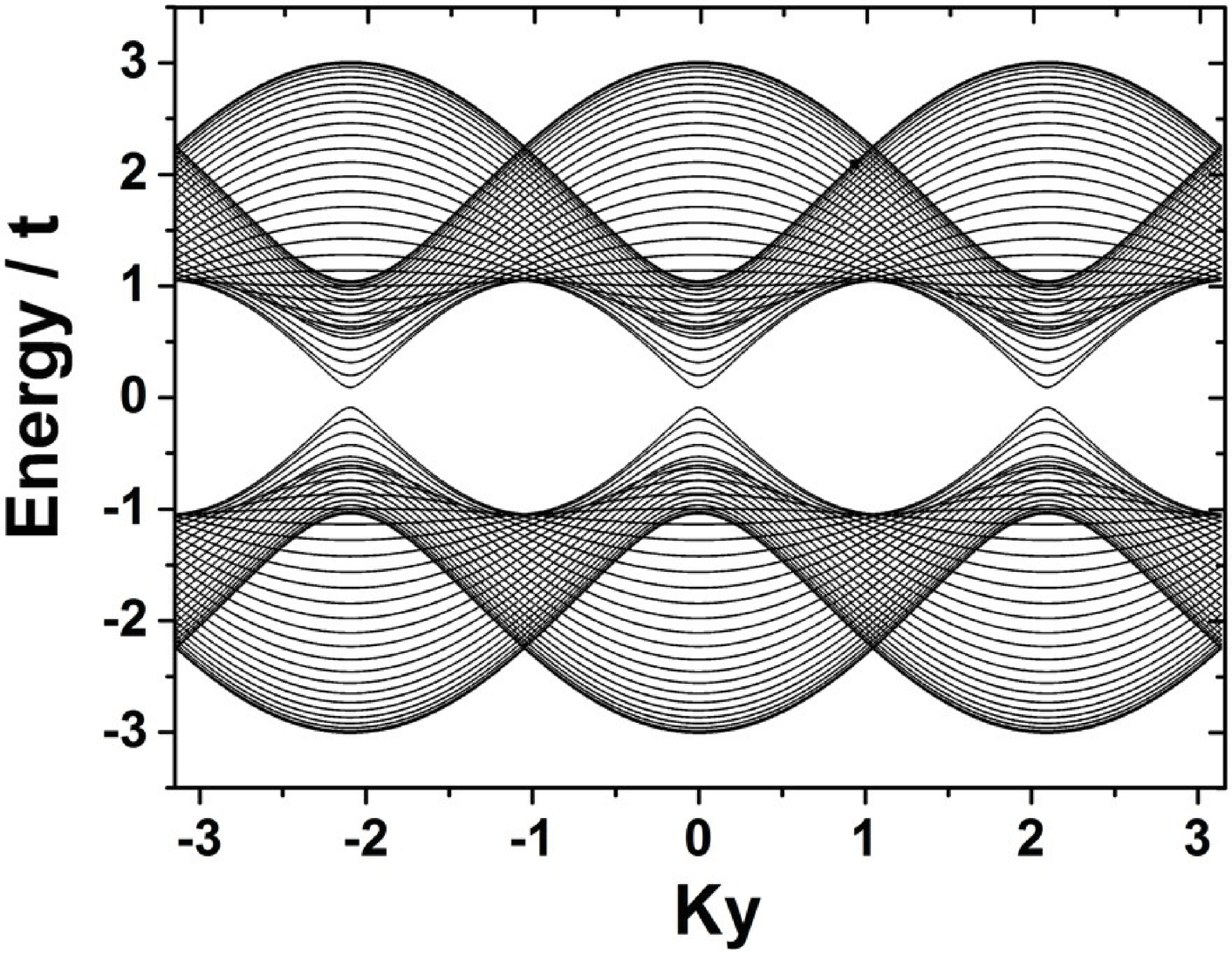}\caption{(Color online) The edge
states of nornal insulator with open armchair boundary. The parameters are
$t^{\prime}/t=0.05$, $\varepsilon/t=0.3$.}%
\label{ni}%
\end{figure}

\begin{figure}[ptbh]
\includegraphics[width = 8.0cm]{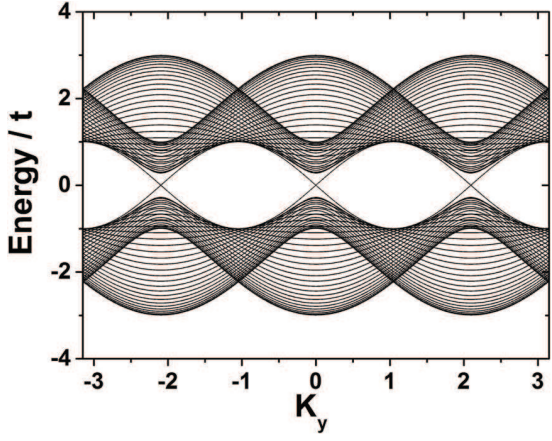}\caption{(Color online) The edge
states of topological insulator (or QAH) with open armchair boundary. The
parameters are $t^{\prime}/t=0.05$, $\varepsilon/t=0.01$.}%
\label{qah}%
\end{figure}

From the spectrum of free fermions, we can see that there exist energy gaps
near the points $\mathbf{k}_{1}=\frac{2\pi}{3}(1,$ $\frac{1}{\sqrt{3}})$ and
$\mathbf{k}_{2}=-\frac{2\pi}{3}(1,$ $\frac{1}{\sqrt{3}})$ as
\[
\Delta E=\left \vert 2\varepsilon-6\sqrt{3}t^{\prime}\right \vert
\]
When the energy gap closed, we can get the phase boundary. We can see that
there exist two phases in the free electron case: the topological insulater
with anomalous quantum Hall effect (we call it QAH) and the normal band
insulator (NI) state. For each spin-component of fermions, the Chern number is
defined by
\begin{equation}
C=\frac{1}{4\pi}\int_{\Omega}d^{2}k[\mathbf{n}\cdot \partial_{x}\mathbf{n}%
\times \partial_{y}\mathbf{n}]=\pm1 \label{c}%
\end{equation}
where $\mathbf{n}$ is defined as $\mathbf{n}=\frac{\mathbf{d}}{|\mathbf{d}|}$
where
\[
\mathbf{d=(}d_{x},d_{y},d_{z}\mathbf{)}%
\]
and
\begin{align*}
d_{x}  &  =\operatorname{Re}\xi_{\mathbf{k}},\\
d_{y}  &  =\operatorname{Im}\xi_{\mathbf{k}},\\
d_{z}  &  =\gamma_{k}+\varepsilon.
\end{align*}
Then at zero temperature ($T=0$), due to spin rotation symmetry, we obtain
$C=\pm2$ in QAH and $C=0$ in NI. The eigen-state of the two states are
depicted in Fig.\ref{ni} and Fig.\ref{qah}.\ One can see that there is gapless
edge states\ in QAH while no edge states in NI.

\section{Mean field phase diagram and topological quantum phase transition}

When increasing the interaction strength, we get an $s$-wave pairing SC order.
The SC order parameter of the $s$-wave SC is
\[
\langle \hat{c}_{i,\uparrow}^{\dagger}\hat{c}_{i,\downarrow}^{\dagger}%
\rangle=\Delta.
\]
Then in the mean field approach, the Hamiltonian can be written as
\begin{equation}
H=H_{\mathrm{H}}+H^{\prime}-U\Delta \sum \limits_{i}\hat{c}_{i,\uparrow
}^{\dagger}\hat{c}_{i,\downarrow}^{\dagger}+h.c..
\end{equation}
In the momentum space, it becomes
\begin{equation}
H=\sum \limits_{k}\Psi_{k}^{\dagger}h_{k}\Psi_{k}%
\end{equation}
where the basis vector $\Psi_{k}^{\dagger}$ is $\left(
\begin{array}
[c]{cccc}%
\hat{a}_{\mathbf{k\uparrow}}^{\dagger} & \hat{a}_{-\mathbf{k\downarrow}} &
\hat{b}_{\mathbf{k\uparrow}}^{\dagger} & \hat{b}_{-\mathbf{k\downarrow}}%
\end{array}
\right)  $ and
\begin{equation}
h_{k}=\left(
\begin{array}
[c]{cccc}%
\gamma_{k}+\varepsilon & -U\Delta & -\xi_{\mathbf{k}} & 0\\
-U\Delta & \gamma_{k}-\varepsilon & 0 & \xi_{\mathbf{k}}\\
-\xi_{\mathbf{k}}^{\ast} & 0 & -\gamma_{k}-\varepsilon & -U\Delta \\
0 & \xi_{\mathbf{k}}^{\ast} & -U\Delta & -\gamma_{k}+\varepsilon
\end{array}
\right)  .
\end{equation}
After diagonalization, we can obtain the spectrum of the quasi-particles
\begin{equation}
\mathbf{E}_{\mathbf{k}_{1}}=\pm \sqrt{(\left \vert \gamma_{k}\right \vert
+\sqrt{\left(  U\Delta \right)  ^{2}+\varepsilon^{2}})^{2}+|\xi_{k}|^{2}}%
\end{equation}
and\
\begin{equation}
\mathbf{E}_{\mathbf{k}_{2}}=\pm \sqrt{(\left \vert \gamma_{k}\right \vert
-\sqrt{\left(  U\Delta \right)  ^{2}+\varepsilon^{2}})^{2}+|\xi_{k}|^{2}}.
\end{equation}

Then in mean field approach, we get the self-consistency equation to derive
$\Delta$ by minimizing the energy in the reduced Brillouin zone as
\begin{align}
1  &  =\frac{1}{4N_{s}}\sum \limits_{\mathbf{k}}[\frac{(\left \vert \gamma
_{k}\right \vert +\sqrt{\left(  U\Delta \right)  ^{2}+\varepsilon^{2}}%
)\frac{\left(  U\right)  }{\sqrt{\left(  U\Delta \right)  ^{2}+\varepsilon^{2}%
}}}{\sqrt{(\left \vert \gamma_{k}\right \vert +\sqrt{\left(  U\Delta \right)
^{2}+\varepsilon^{2}})^{2}+|\xi_{k}|^{2}}}\nonumber \\
&  -\frac{(\left \vert \gamma_{k}\right \vert -\sqrt{\left(  U\Delta \right)
^{2}+\varepsilon^{2}})\frac{\left(  U\right)  }{\sqrt{\left(  U\Delta \right)
^{2}+\varepsilon^{2}}}}{\sqrt{(\left \vert \gamma_{k}\right \vert -\sqrt{\left(
U\Delta \right)  ^{2}+\varepsilon^{2}})^{2}+|\xi_{k}|^{2}}}].
\end{align}

\begin{figure}[ptbh]
\includegraphics[width = 9.0cm]{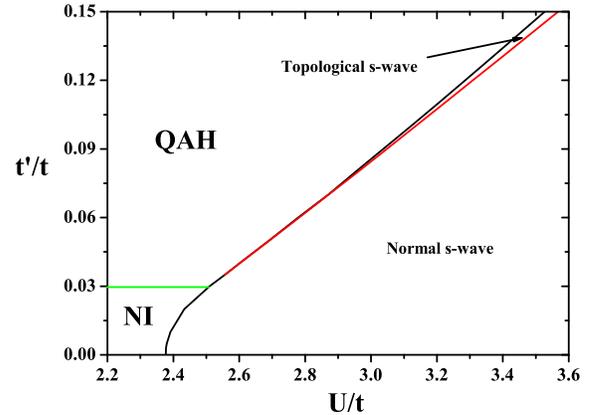}\caption{(Color online) The phase
diagram of the case $\varepsilon/t=0.15\ $at $T=0.$ There exist four phases :
the NI state, the QAH state, the topological s-wave state and the normal
s-wave. }%
\label{phase}%
\end{figure}

To determine the phase diagram, there are four types of phase transitions :
the quantum phase transition between\ topological $s$-wave SC order with
$\Delta \neq0$ and QAH with $\Delta=0$, the topological quantum phase
transition between\ topological $s$-wave SC order with $\Delta \neq0$ and the
normal $s$-wave SC state with $\Delta \neq0,$ the quantum phase transition
between NI with $\Delta=0$ and QAH with $\Delta=0$, the quantum phase
transition between NI and the normal $s$-wave SC state with $\Delta \neq0.$ In
particular, the topological quantum phase transition between TSC and normal
$s$-wave SC is determined by the condition of zero quasi-particle's energy
gap
\begin{equation}
\Delta_{f}=2(3\sqrt{3}t^{\prime}-\sqrt{\left(  U\Delta \right)  ^{2}%
+\varepsilon^{2}})=0.
\end{equation}
After determining the phase boundaries, we plot the phase diagram in
Fig.\ref{phase}. In the phase diagram of Fig.\ref{phase}, there exist four
different quantum phases : QAH, NI, $C=\pm2$\ topological $s$-wave SC and the
normal $s$-wave SC.

From Fig.\ref{phase}, one can see that in the non-interacting limit $\left(
U=0\right)  ,$ the ground state is a $C=\pm2$ topological insulator with QAH
for $t^{\prime}>0.0288t$ and NI for $t^{\prime}<0.0288t$. At $t^{\prime
}=0.0288t,$\ the electron energy gap closes at high symmetry points in the
momentum space. As a result, a third order topological quantum phase
transition occurs between QAH and NI. See the dispersion of electrons for
$t^{\prime}=0.0288t$ in Fig.5. With the increasing of the on-site Coulomb
interaction strength, the ground state can be an $s$-wave superconductor. For
$t^{\prime}>0.0288t,$ the quantum phase transition between QAH and $s$-wave
order is always first order which is denoted by the black line in
Fig.\ref{phase}.\begin{figure}[ptb]
\includegraphics[width=0.4\textwidth]{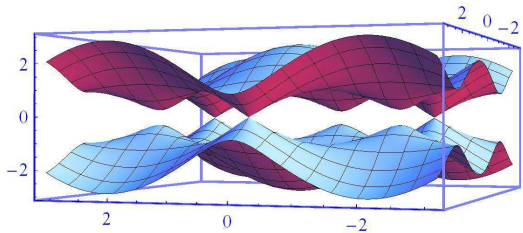}\caption{(Color online) The
dispersion of electrons for $t^{\prime}=0.0288t$ when $U=0.$ We can see
clearly that in the high symmetry point the energy gap is zero and the
dispersion has a Dirac cone.}%
\end{figure}

\begin{figure}[ptbh]
\includegraphics[width = 8.0cm]{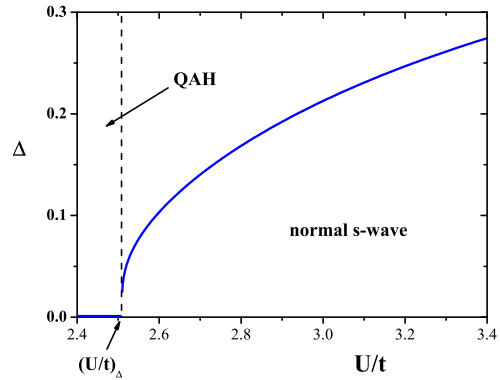}\caption{(Color online) The
superconductor order $\Delta$ for the case of $\varepsilon/t=0.15$ and
$t^{\prime}/t=0.03.$ $(\frac{U}{t})_{\Delta}$ is the critical point of the
superconductor order phase transition.}%
\label{cri1}%
\end{figure}

\begin{figure}[ptbh]
\includegraphics[width = 8.0cm]{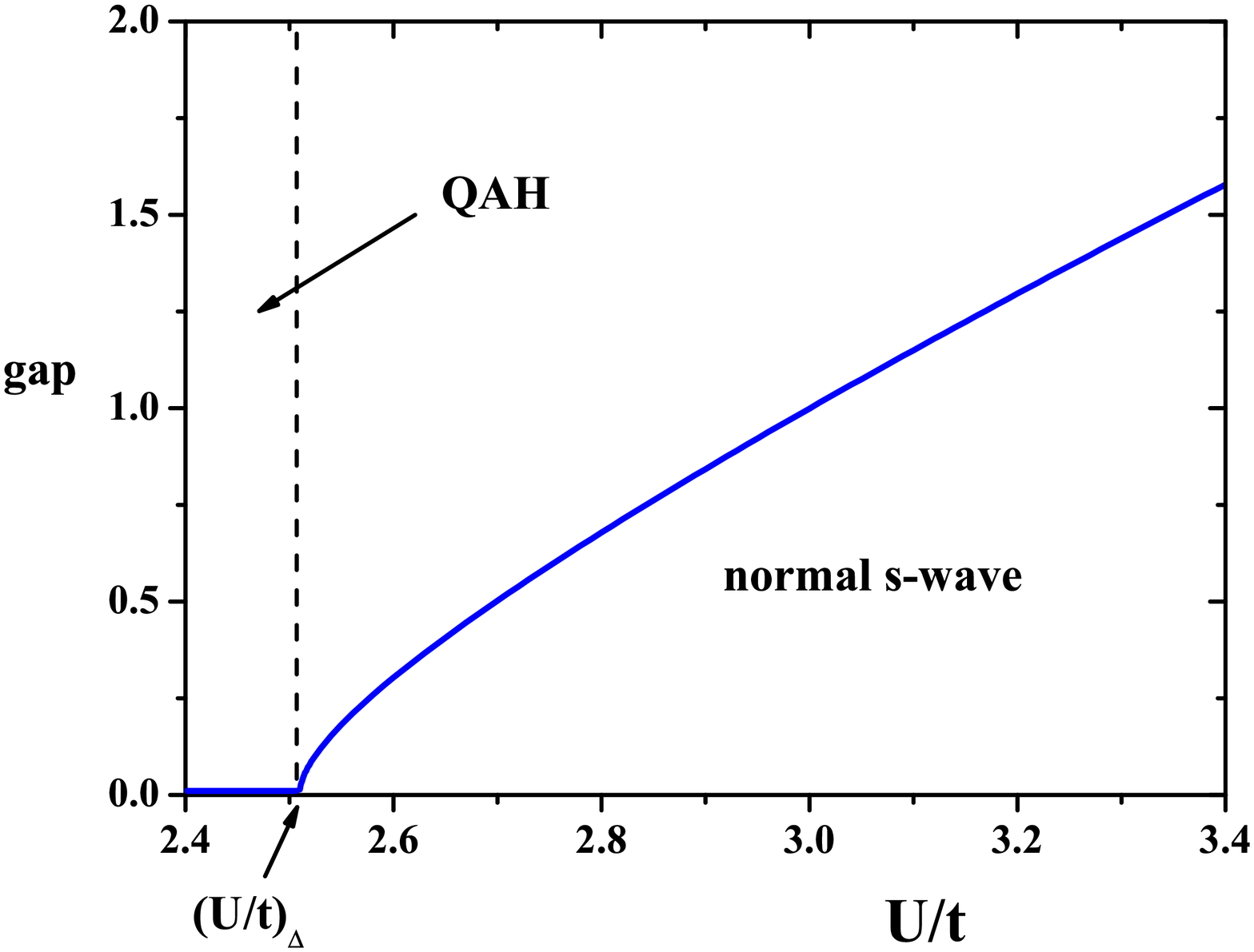}\caption{(Color online) The energy
gap $\Delta_{f}$ for the case of $\varepsilon/t=0.15$ and $t^{\prime}/t=0.03.$
$(\frac{U}{t})_{\Delta}$ is the critical point of the superconductor order
phase transition.}%
\label{cri2}%
\end{figure}

\begin{figure}[ptbh]
\includegraphics[width = 8.0cm]{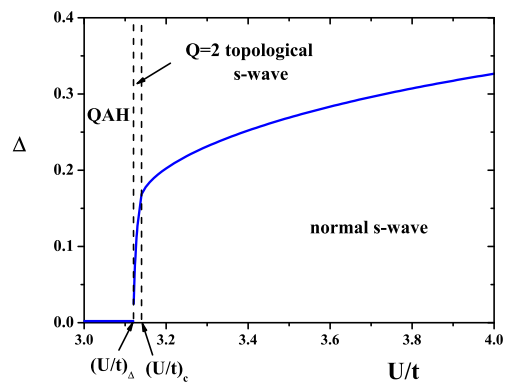}\caption{(Color online) The
superconductor order $\Delta$ for the case of $\varepsilon/t=0.15$ and
$t^{\prime}/t=0.1.$ $(\frac{U}{t})_{\Delta}$ is the critical point of the
superconductor order phase transition and $(\frac{U}{t})_{c}$ is the
topological quantum phase transition.}%
\label{cri3}%
\end{figure}

\begin{figure}[ptbh]
\includegraphics[width = 8.0cm]{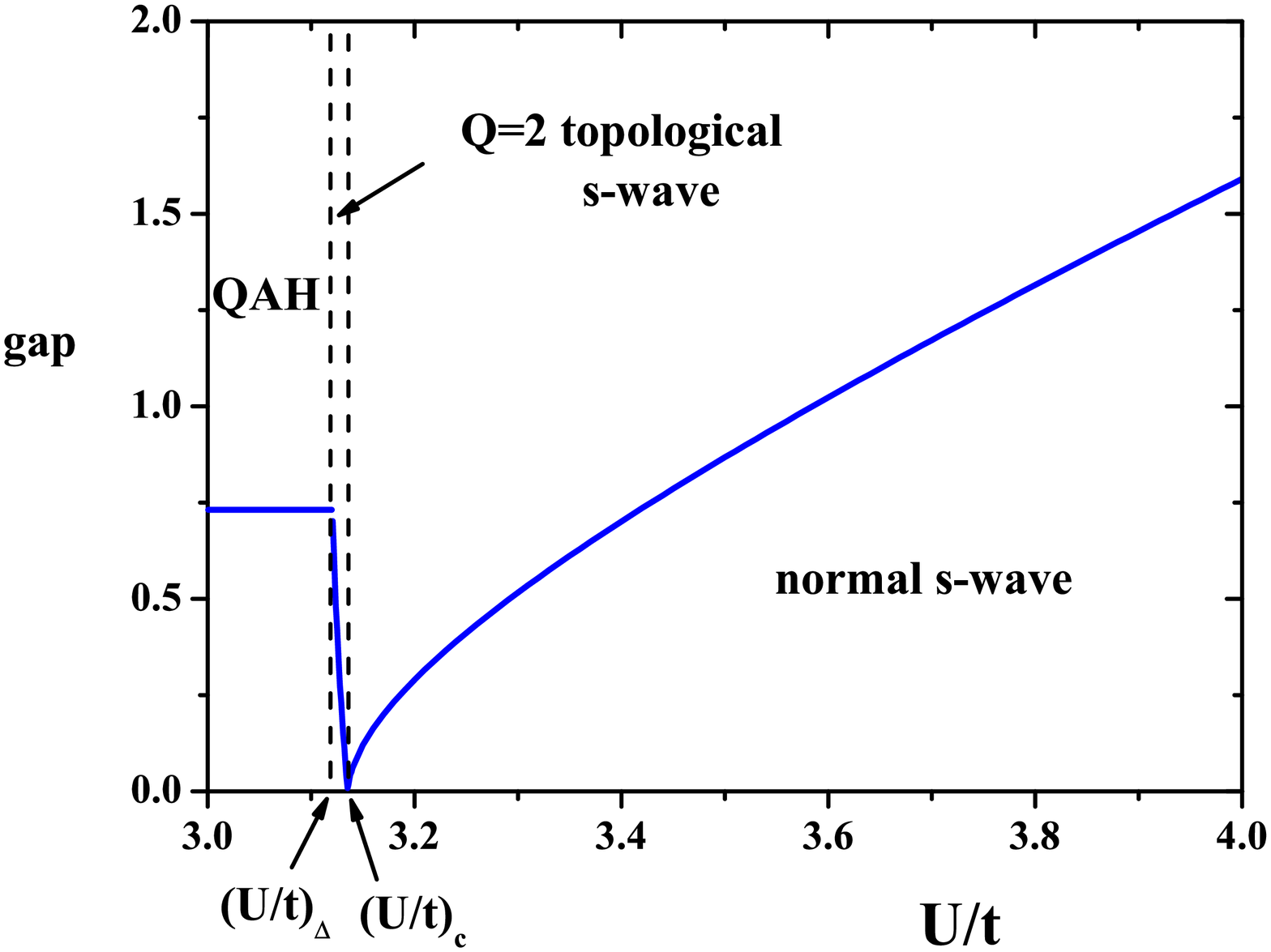}\caption{(Color online) The energy
gap $\Delta_{f}$ for the case of $\varepsilon/t=0.15$ and $t^{\prime}/t=0.1.$
$(\frac{U}{t})_{\Delta}$ is the critical point of the superconductor order
phase transition and $(\frac{U}{t})_{c}$ is the topological quantum phase
transition.}%
\label{cri4}%
\end{figure}

\begin{figure}[ptbh]
\includegraphics[width = 8.0cm]{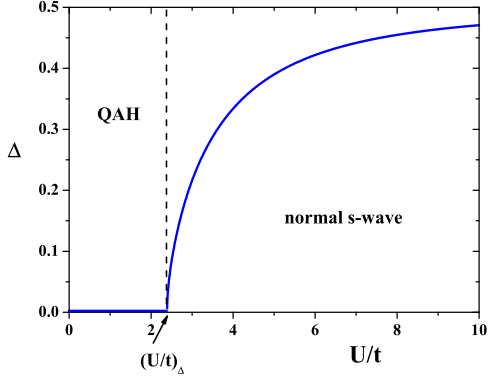}\caption{(Color online) The
superconductor order $\Delta$ for the case of $\varepsilon/t=0.15$ and
$t^{\prime}/t=0.01.$ $(\frac{U}{t})_{\Delta}$ is the critical point of the
superconductor order phase transition.}%
\label{cri5}%
\end{figure}

In the region of $0.0288t<t^{\prime}<0.035t$, due to the jumping of the
superconductor order, the QAH state will turn into normal $s$-wave state
without gap closing. In Fig.\ref{cri1} and Fig.\ref{cri2}, we plot the SC
order parameter\ and the energy gap for the case of $\varepsilon/t=0.15,$
$t^{\prime}/t=0.03$. While in the region of $t^{\prime}>0.035t,$ with the
increasing of the interaction strength, the QAH state will turn into the
topological $s$-wave SC order after crossing a first order phase transition
(black line in Fig.\ref{phase}) and then turn into the normal $s$-wave order
crossing a second order quantum phase transition (red line in Fig.\ref{phase}%
). In Fig.\ref{cri3} and Fig.\ref{cri4}, we also plot the SC order\ parameter
and the energy gap for the case of $\varepsilon/t=0.15,$ $t^{\prime}/t=0.1$.
In the region of $t^{\prime}<0.0288t,$ the quantum phase transition between NI
and $s$-wave SC order is always second order which is denoted by the black
line in Fig.\ref{phase}. With the increase of the\ interaction strength $U,$
due to the smoothly change of the SC order parameter, the NI state will turn
into the normal $s$-wave order after crossing a second order phase transition.
In Fig.\ref{cri5}, we plot the SC order parameter for the case of
$\varepsilon/t=0.15,$ $t^{\prime}/t=0.01$.

\section{Topological properties of TSC}

In this section we will study the topological properties in the TSC state. The
$s$-wave paired ground state exhibits a non-trivial topological property that
can be signified by the Chern number $C=\pm2$ in the momentum space. Because
the pairing is s-wave, such TSC\ belongs to topological superconductor. In
this section we calculate the edge states and the zero modes around the $\pi$-flux.

\subsection{Edge states}

In this part, we study the edge states of TSC. The dispersion of the edge
states of normal SC and that of TSC are plotted in Fig.11 and Fig.12,
respectively. From them, we find there exist no gapless edge states in the
normal SC. While for TSC there always exist two gapless edge states.

\begin{figure}[ptbh]
\includegraphics[width = 8.0cm]{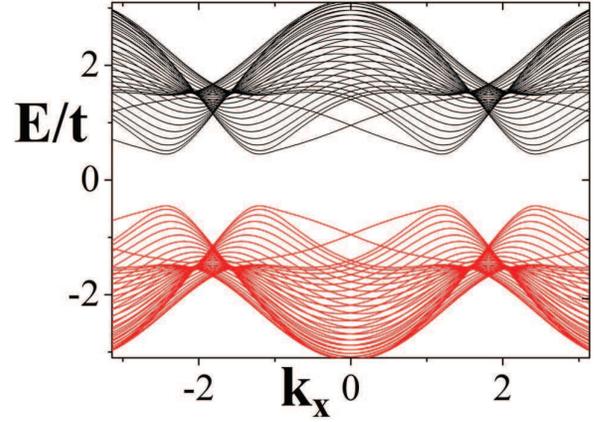}\caption{(Color online) The edge
states of normal SC with "zigzag" open boundary for parameters $t^{\prime
}/t=0.1$, $\varepsilon/t=0.15$, $U/t=3.5$.}%
\label{nsc}%
\end{figure}\begin{figure}[ptbh]
\includegraphics[width = 8.0cm]{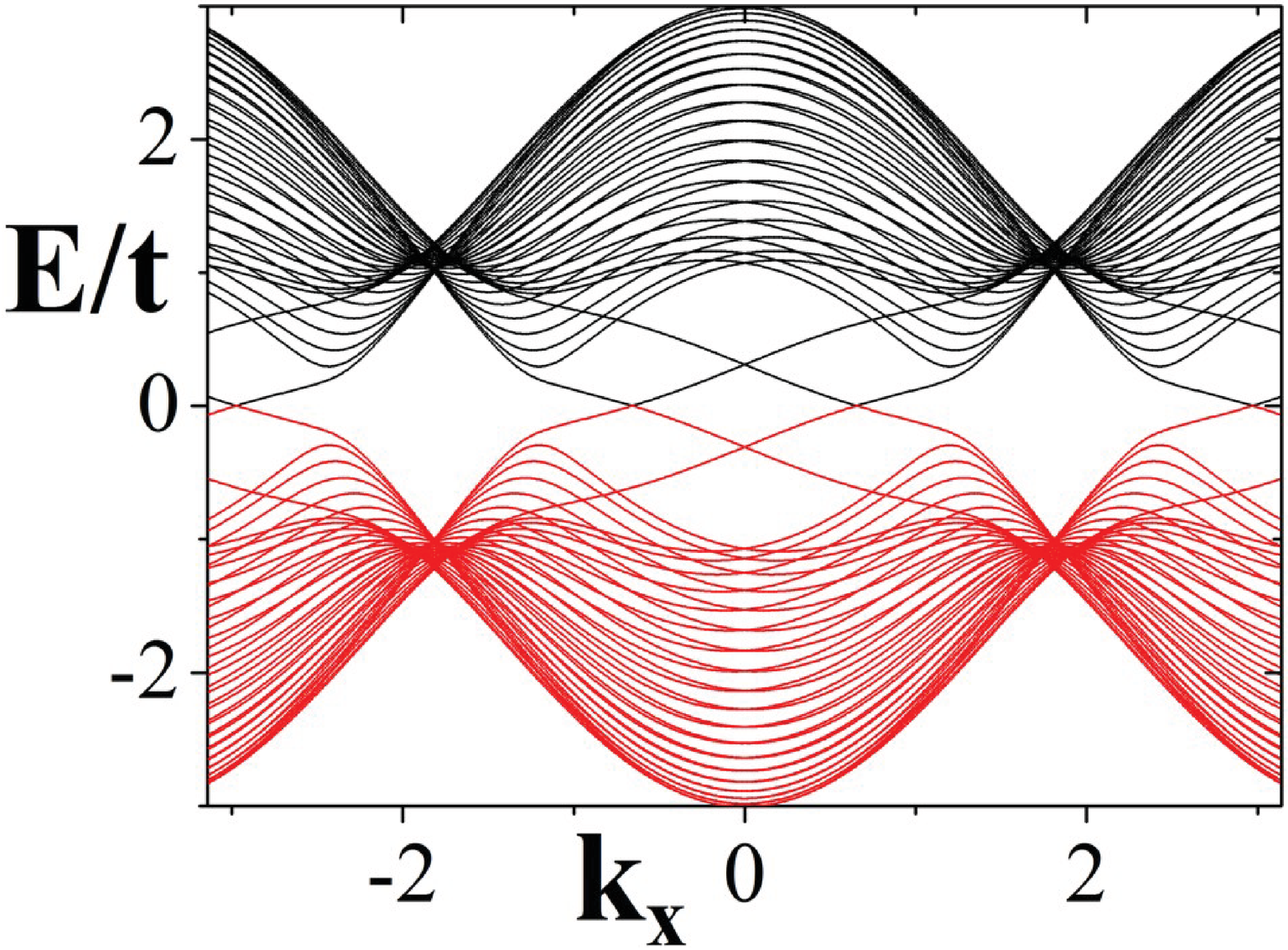}\caption{(Color online) The edge
states of TSC with "zigzag" open boundary for parameters $t^{\prime}/t=0.1$,
$\varepsilon/t=0.15$, $U/t=3.125$.}%
\label{TSC1}%
\end{figure}

\subsection{Zero modes of $\pi$-flux}

\begin{figure}[ptb]
\includegraphics[width=0.5\textwidth]{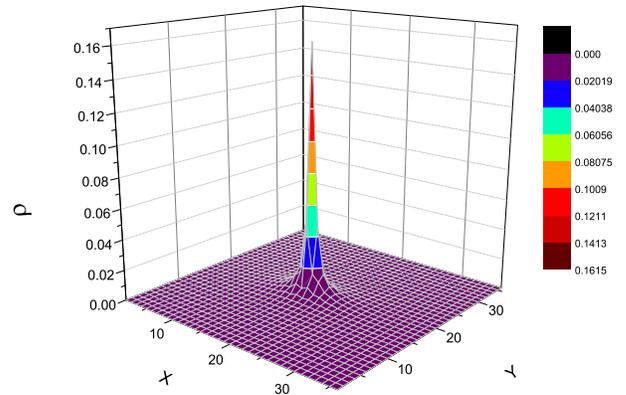}\caption{(Color online) The
particle density around a $\pi$-flux in TSC.}%
\label{zero}%
\end{figure}

After recognizing the properties of edge states, we turn to study $\pi$-flux.
A $\pi$-flux denotes half a flux quantum one plaquette of the square lattice,
$\frac{1}{2}\Phi_{0}$ $(\Phi_{0}=\frac{hc}{e})$. For a TSC on a honeycomb
lattice, a $\pi$-flux on a plaquette is confined at zero temperature and
cannot be real excitation. From the numerical calculations, we found that
there exist two zero modes around each $\pi$-flux. And in Fig.13, we plot the
particle density around a $\pi$-flux in TSC phase.

\section{Phase stiffness}

In this section we study the phase fluctuations of TSC going beyond mean field
method by random-phase-approximation (RPA) approach \cite{flu}. Now the SC
order parameters become $\Delta(i)=\left \vert \Delta \right \vert e^{i\theta
_{i}}$. For the fluctuated SC order parameters $\Delta(i),$ there are two
bosonic modes : a gapped amplitude mode and the Goldstone mode (the mode
describing phase fluctuations).

In the imaginary-time path-integral representation ($\beta=1/T$,
$\hslash=k_{B}=1$), using the Hubbard-Stratonovich transformation, the
partition function can be written as\cite{tay}%
\begin{equation}
Z=\int[dc_{\sigma}^{\ast}dc_{\sigma}][d\Delta^{\ast}d\Delta]e^{-S(c_{\sigma
}^{\ast},c_{\sigma},\Delta^{\ast},\Delta)}%
\end{equation}%
\begin{align*}
S  &  =\int_{0}^{\beta}d\tau \sum_{\mathbf{r}}\{ \frac{1}{U}\Delta^{\ast}%
\Delta+c_{\sigma}^{\ast}\partial_{\tau}c_{\sigma}^{\ast}+H_{t}+H_{t^{\prime}%
}+H_{\varepsilon}\\
&  -\Delta(\mathbf{r},\tau)c_{\sigma}(\mathbf{r},\tau)c_{\bar{\sigma}%
}(\mathbf{r},\tau)-\Delta^{\ast}(\mathbf{r},\tau)c_{\bar{\sigma}}^{\ast
}(\mathbf{r},\tau)c_{\sigma}^{\ast}(\mathbf{r},\tau)\}.
\end{align*}
After integrating over the electron fields, in the momentum space, the action
can be written as
\[
S=\frac{\beta}{U}\sum \nolimits_{q,\eta=A,B}\Delta_{\eta}^{\ast}(q)\Delta
_{\eta}(q)+2\beta \mu N-\sum \nolimits_{q}\ln G^{-1}%
\]
where $\Delta_{\eta}(q)=\Delta_{\eta,0\text{ }}+\Lambda_{\eta}(q)$,
$\Delta_{\eta,0\text{ }}=\Delta_{0\text{ }}$ and $G^{-1}=G_{0}^{-1}+G_{1}%
^{-1}$ with \begin{widetext}
\[
G_{0}^{-1}=\left(
\begin{array}
[c]{cccc}%
\mathrm{i}\omega_{m}+\gamma_{\mathbf{k}}+\varepsilon & -\Delta_{0\text{ }} &
-\xi_{k} & 0\\
-\Delta_{0\text{ }} & \mathrm{i}\omega_{m}+\gamma_{\mathbf{k}}-\varepsilon &
0 & \xi_{k}\\
-\xi_{k}^{\ast} & 0 & \mathrm{i}\omega_{m}-\gamma_{\mathbf{k}}-\varepsilon &
-\Delta_{0\text{ }}\\
0 & \xi_{k}^{\ast} & -\Delta_{0\text{ }} & \mathrm{i}\omega_{m}-\gamma
_{\mathbf{k}}+\varepsilon
\end{array}
\right)
\]
\end{widetext}and%
\begin{equation}
G_{1}^{-1}=\left(
\begin{array}
[c]{cccc}%
0 & -\Lambda_{A}(q) & 0 & 0\\
-\Lambda_{A}^{\ast}(-q) & 0 & 0 & 0\\
0 & 0 & 0 & -\Lambda_{B}(q)\\
0 & 0 & -\Lambda_{B}^{\ast}(-q) & 0
\end{array}
\right)  .
\end{equation}
Here we have used
\begin{align*}
\Lambda_{\eta}(q)  &  =\frac{1}{\sqrt{N}}\int dx\Lambda_{\eta}%
(x)e^{-\mathrm{i}q.x}\\
\Lambda_{\eta}^{\ast}(-q)  &  =\frac{1}{\sqrt{N}}\int dx\Lambda_{\eta}^{\ast
}(x)e^{-\mathrm{i}q.x}\\
\Lambda_{\eta}(x)  &  =\frac{1}{\sqrt{N}}\sum_{q}\Lambda_{\eta}%
(q)e^{\mathrm{i}q.x}\\
\Lambda_{\eta}^{\ast}(x)  &  =\frac{1}{\sqrt{N}}\sum_{q}\Lambda_{\eta}^{\ast
}(q)e^{-\mathrm{i}q.x}%
\end{align*}

Next, we investigate the Gaussian fluctuations around the saddle point. Using
the expansion of the natural logarithm of
\begin{align}
\text{\textrm{Tr}}\ln G^{-1}  &  =\ln(G_{0}^{-1}+G_{1}^{-1})\\
&  =\text{\textrm{Tr}}\ln G_{0}^{-1}-\sum_{l=1}^{\infty}\frac{(-1)^{l}}%
{l}\text{\textrm{Tr}}(G_{0}G_{1}^{-1})^{l},\nonumber
\end{align}
we expand the action to second order of the fluctuation fields $\Lambda_{\eta
}(q)$. This procedure leads to an effective action as
\begin{equation}
S=S_{0}+S_{1}^{\prime}%
\end{equation}
where the effective action at the saddle point is%
\begin{equation}
S_{0}=\frac{\beta}{U}\sum \nolimits_{\eta=A,B}\Delta_{\eta,0}\Delta_{\eta
,0}+2\beta \mu N-\text{\textrm{Tr}}\ln G_{0}^{-1}%
\end{equation}
and the effective action corresponding the quantum fluctuations reads%
\begin{align}
S_{1}^{\prime}  &  =\frac{1}{2}\text{\textrm{Tr}}\sum_{q,p}(G_{0}(p)G_{1}%
^{-1}(p,p-q)G_{0}(p-q)G_{1}^{-1}(p-q,p))\nonumber \\
&  +\frac{\beta}{U}\sum \nolimits_{q,\eta=A,B}\Delta_{\eta}^{\ast}%
(q)\Delta_{\eta}(q)\nonumber \\
&  =S_{1}+S_{2}%
\end{align}
where
\begin{equation}
S_{2}=\frac{\beta}{U}\sum \nolimits_{q,\eta=A,B}\Delta_{\eta}^{\ast}%
(q)\Delta_{\eta}(q)
\end{equation}
and
\begin{equation}
S_{1}=\frac{\beta}{2}\sum_{q}\Lambda_{q}^{\dag}Q^{\prime}(q)\Lambda_{q}.
\end{equation}
Here $\Lambda_{q}^{\dag}=(\Lambda_{A}^{\ast}(-q),\Lambda_{A}(q),\Lambda
_{B}(q),\Lambda_{B}^{\ast}(-q))$ denotes the fluctuation field and $Q^{\prime
}(q)$ is given by the following matrix
\begin{equation}
Q^{\prime}(q)=\left(
\begin{array}
[c]{cccc}%
Q_{11}^{\prime}(q) & Q_{12}^{\prime}(q) & Q_{13}^{\prime}(q) & Q_{14}^{\prime
}(q)\\
Q_{21}^{\prime}(q) & Q_{22}^{\prime}(q) & Q_{23}^{\prime}(q) & Q_{24}^{\prime
}(q)\\
Q_{31}^{\prime}(q) & Q_{32}^{\prime}(q) & Q_{33}^{\prime}(q) & Q_{34}^{\prime
}(q)\\
Q_{41}^{\prime}(q) & Q_{42}^{\prime}(q) & Q_{43}^{\prime}(q) & Q_{44}^{\prime
}(q)
\end{array}
\right)
\end{equation}
where%
\begin{align}
Q_{11}^{\prime}(q)  &  =\frac{1}{\beta N}\sum_{p}G_{022}(p)G_{011}%
(p-q),\nonumber \\
Q_{12}^{\prime}(q)  &  =\frac{1}{\beta N}\sum_{p}G_{012}(p)G_{012}%
(p-q),\nonumber \\
Q_{13}^{\prime}(q)  &  =\frac{1}{\beta N}\sum_{p}G_{032}(p)G_{014}%
(p-q),\nonumber \\
Q_{14}^{\prime}(q)  &  =\frac{1}{\beta N}\sum_{p}G_{042}(p)G_{013}(p-q),
\end{align}

\begin{align}
Q_{21}^{\prime}(q)  &  =\frac{1}{\beta N}\sum_{p}G_{021}(p)G_{021}%
(p-q),\nonumber \\
Q_{22}^{\prime}(q)  &  =\frac{1}{\beta N}\sum_{p}G_{011}(p)G_{022}%
(p-q),\nonumber \\
Q_{23}^{\prime}(q)  &  =\frac{1}{\beta N}\sum_{p}G_{031}(p)G_{024}%
(p-q),\nonumber \\
Q_{24}^{\prime}(q)  &  =\frac{1}{\beta N}\sum_{p}G_{041}(p)G_{023}(p-q),
\end{align}%
\begin{align}
Q_{31}^{\prime}(q)  &  =\frac{1}{\beta N}\sum_{p}G_{023}(p)G_{041}%
(p-q),\nonumber \\
Q_{32}^{\prime}(q)  &  =\frac{1}{\beta N}\sum_{p}G_{013}(p)G_{042}%
(p-q),\nonumber \\
Q_{33}^{\prime}(q)  &  =\frac{1}{\beta N}\sum_{p}G_{033}(p)G_{044}%
(p-q),\nonumber \\
Q_{34}^{\prime}(q)  &  =\frac{1}{\beta N}\sum_{p}G_{043}(p)G_{043}(p-q),
\end{align}%
\begin{align}
Q_{41}^{\prime}(q)  &  =\frac{1}{\beta N}\sum_{p}G_{024}(p)G_{031}%
(p-q),\nonumber \\
Q_{42}^{\prime}(q)  &  =\frac{1}{\beta N}\sum_{p}G_{014}(p)G_{032}%
(p-q),\nonumber \\
Q_{43}^{\prime}(q)  &  =\frac{1}{\beta N}\sum_{p}G_{034}(p)G_{034}%
(p-q),\nonumber \\
Q_{44}^{\prime}(q)  &  =\frac{1}{\beta N}\sum_{p}G_{044}(p)G_{033}(p-q).
\end{align}
See the details of $G_{0}$ in appendix.

For the term $Q_{ij}^{\prime}(q)$, using the Matsubara summation formula as
$1/\beta \sum_{\omega_{m}}1/(\mathrm{i}\omega_{m}-E)=n_{F}(E)$, at zero
temperature we may obtain\begin{widetext}
\begin{align*}
Q_{ij}^{\prime}(q)  & =\frac{1}{\beta N}\sum_{p}G_{0kl}(p)G_{0mn}(p-q)\\
& =\frac{1}{N}\sum_{\mathbf{k}}\frac{1}{\beta}\sum_{\omega_{m}}(\frac{A_{kl}%
}{\mathrm{i}\omega_{m}+E_{+}(\mathbf{k})}+\frac{B_{kl}}{\mathrm{i}\omega
_{m}-E_{+}(\mathbf{k})}+\\
& \frac{C_{kl}}{\mathrm{i}\omega_{m}+E_{-}(\mathbf{k})}+\frac{D_{kl}%
}{\mathrm{i}\omega_{m}-E_{-}(\mathbf{k})})(\frac{A_{mn}}{\mathrm{i}\omega
_{m}-\mathrm{i}\omega_{n}+E_{+}(\mathbf{k-q})}\\
& +\frac{B_{mn}}{\mathrm{i}\omega_{m}-\mathrm{i}\omega_{n}-E_{+}%
(\mathbf{k-q})}+\frac{C_{mn}}{\mathrm{i}\omega_{m}-\mathrm{i}\omega_{n}%
+E_{-}(\mathbf{k-q})}\\
& +\frac{D_{mn}}{\mathrm{i}\omega_{m}-\mathrm{i}\omega_{n}-E_{-}%
(\mathbf{k-q})})\\
& =\frac{1}{N}[\sum_{\mathbf{k}}\frac{A_{kl}B_{mn}}{-\mathrm{i}\omega
_{n}-E_{+}(\mathbf{k-q})-E_{+}(\mathbf{k})}+\sum_{\mathbf{k}}\frac
{A_{kl}D_{mn}}{-\mathrm{i}\omega_{n}-E_{-}(\mathbf{k-q})-E_{+}(\mathbf{k})}\\
& \sum_{\mathbf{k}}\frac{-B_{kl}A_{mn}}{-\mathrm{i}\omega_{n}+E_{+}%
(\mathbf{k-q})+E_{+}(\mathbf{k})}+\sum_{\mathbf{k}}\frac{-B_{kl}C_{mn}%
}{-\mathrm{i}\omega_{n}+E_{-}(\mathbf{k-q})+E_{+}(\mathbf{k})}\\
& \sum_{\mathbf{k}}\frac{C_{kl}B_{mn}}{-\mathrm{i}\omega_{n}-E_{+}%
(\mathbf{k-q})-E_{-}(\mathbf{k})}+\sum_{\mathbf{k}}\frac{C_{kl}D_{mn}%
}{-\mathrm{i}\omega_{n}-E_{-}(\mathbf{k-q})-E_{-}(\mathbf{k})}\\
& \sum_{\mathbf{k}}\frac{-D_{kl}A_{mn}}{-\mathrm{i}\omega_{n}+E_{+}%
(\mathbf{k-q})+E_{-}(\mathbf{k})}+\sum_{\mathbf{k}}\frac{D_{kl}C_{mn}%
}{-\mathrm{i}\omega_{n}+E_{-}(\mathbf{k-q})+E_{-}(\mathbf{k})}]
\end{align*}
\end{widetext}where $\omega_{m}=(2n+1)\pi/\beta$, $\omega_{n}=2n\pi/\beta$,
and $n_{F}(E)$ is the Fermi distribution function which reads $n_{F}%
(E)=1/(e^{\beta E}+1)$. When the usual analytic continuation $\mathrm{i}%
\omega_{n}\rightarrow \omega+\mathrm{i}0^{+}$ is performed, and in static limit
$\omega=0$, $Q_{ij}^{\prime}(q)$ is reduced into%
\begin{align}
Q_{ij}^{\prime}(q)  &  =\frac{-1}{N}[\sum_{\mathbf{k}}\frac{A_{kl}%
(\mathbf{k})B_{mn}(\mathbf{k-q})+B_{kl}(\mathbf{k})A_{mn}(\mathbf{k}%
-\mathbf{q})}{E_{+}(\mathbf{k-q})+E_{+}(\mathbf{k})}\nonumber \\
&  +\sum_{\mathbf{k}}\frac{A_{kl}(\mathbf{k})D_{mn}(\mathbf{k-q}%
)+B_{kl}(\mathbf{k})C_{mn}(\mathbf{k-q})}{E_{-}(\mathbf{k-q})+E_{+}%
(\mathbf{k})}+\nonumber \\
&  \sum_{\mathbf{k}}\frac{C_{kl}(\mathbf{k})B_{mn}(\mathbf{k-q})+D_{kl}%
(\mathbf{k})A_{mn}(\mathbf{k-q})}{E_{+}(\mathbf{k-q})+E_{-}(\mathbf{k}%
)}+\nonumber \\
&  \sum_{\mathbf{k}}\frac{C_{kl}(\mathbf{k})D_{mn}(\mathbf{k-q})+D_{kl}%
(\mathbf{k})C_{mn}(\mathbf{k-q})}{E_{-}(\mathbf{k-q})+E_{-}(\mathbf{k})}].
\end{align}

In order to obtain the phase stiffness, we can express $\Lambda_{\eta}%
(q)=\chi_{\eta}(q)e^{\mathrm{i}{\normalsize \varphi}_{{\normalsize \eta}%
}{\normalsize (q)}}=[\lambda_{\eta}(q)+\mathrm{i}\theta_{\eta}(q)]/\sqrt{2}$,
where $\chi_{\eta}(q)$, ${\normalsize \varphi}_{{\normalsize \eta}%
}{\normalsize (q)}$, $\lambda_{\eta}(q)$ and $\theta_{\eta}(q)$ are the real
fields. $\lambda_{\eta}(q)$ and $\theta_{\eta}(q)$ essentially can be regarded
as the amplitude field and the phase field, respectively. For the general
cases, ${\normalsize \varphi}_{{\normalsize \eta}}{\normalsize (q)}$ is
small\cite{flu}. Then the field $\Lambda_{q}$ in the new basis $\left(
\begin{array}
[c]{c}%
\lambda_{A}(q)\\
\theta_{A}(q)\\
\theta_{B}(q)\\
\lambda_{B}(q)
\end{array}
\right)  $ is given by%
\begin{equation}
\Lambda_{q}=\frac{1}{\sqrt{2}}\left(
\begin{array}
[c]{cccc}%
1 & \mathrm{i} & 0 & 0\\
1 & -\mathrm{i} & 0 & 0\\
0 & 0 & -\mathrm{i} & 1\\
0 & 0 & \mathrm{i} & 1
\end{array}
\right)  .
\end{equation}

In a rotated basis, the matrix $Q^{\prime}(q)$ can be changed into $\tilde
{Q}(q)$. Now we have the action of the quantum fluctuations as
\begin{equation}
S_{1}=\frac{\beta}{2}\sum_{q}\Phi_{q}^{\dag}\tilde{Q}(q)\Phi_{q}%
\end{equation}
where $\Phi_{q}^{\dag}=(\lambda_{A}(q),\theta_{A}(q),\theta_{B}(q),\lambda
_{B}(q)$ and $\tilde{Q}(q)$ in this rotated basis reads%
\begin{align}
\tilde{Q}_{11}(q)  &  =(Q_{11}^{\prime}+Q_{21}^{\prime}+Q_{12}^{\prime}%
+Q_{22}^{\prime})/2,\nonumber \\
\tilde{Q}_{12}(q)  &  =(\mathrm{i}Q_{11}^{\prime}+\mathrm{i}Q_{21}^{\prime
}-\mathrm{i}Q_{12}^{\prime}-\mathrm{i}Q_{22}^{\prime})/2,\nonumber \\
\tilde{Q}_{13}(q)  &  =(-\mathrm{i}Q_{13}^{\prime}-\mathrm{i}Q_{23}^{\prime
}+\mathrm{i}Q_{14}^{\prime}+\mathrm{i}Q_{24}^{\prime})/2,\nonumber \\
\tilde{Q}_{14}(q)  &  =(Q_{13}^{\prime}+Q_{23}^{\prime}+Q_{14}^{\prime}%
+Q_{24}^{\prime})/2,\nonumber \\
\tilde{Q}_{21}(q)  &  =(-\mathrm{i}Q_{11}^{\prime}+\mathrm{i}Q_{21}^{\prime
}-\mathrm{i}Q_{12}^{\prime}+\mathrm{i}Q_{22}^{\prime})/2,\nonumber \\
\tilde{Q}_{22}(q)  &  =(Q_{11}^{\prime}-Q_{21}^{\prime}-Q_{12}^{\prime}%
+Q_{22}^{\prime})/2,\nonumber \\
\tilde{Q}_{23}(q)  &  =(-Q_{13}^{\prime}+Q_{23}^{\prime}+Q_{14}^{\prime
}-Q_{24}^{\prime})/2,\nonumber \\
\tilde{Q}_{24}(q)  &  =(-\mathrm{i}Q_{13}^{\prime}+\mathrm{i}Q_{23}^{\prime
}-\mathrm{i}Q_{14}^{\prime}+\mathrm{i}Q_{24}^{\prime})/2,\nonumber \\
\tilde{Q}_{31}(q)  &  =(\mathrm{i}Q_{31}^{\prime}-\mathrm{i}Q_{41}^{\prime
}+\mathrm{i}Q_{32}^{\prime}-\mathrm{i}Q_{42}^{\prime})/2,\nonumber \\
\tilde{Q}_{32}(q)  &  =(-Q_{31}^{\prime}+Q_{41}^{\prime}+Q_{32}^{\prime
}-Q_{42}^{\prime})/2,\nonumber \\
\tilde{Q}_{33}(q)  &  =(Q_{33}^{\prime}-Q_{43}^{\prime}-Q_{34}^{\prime}%
+Q_{44}^{\prime})/2,\nonumber \\
\tilde{Q}_{34}(q)  &  =(\mathrm{i}Q_{33}^{\prime}-\mathrm{i}Q_{43}^{\prime
}+\mathrm{i}Q_{34}^{\prime}-\mathrm{i}Q_{44}^{\prime})/2,\nonumber \\
\tilde{Q}_{41}(q)  &  =(Q_{31}^{\prime}+Q_{41}^{\prime}+Q_{32}^{\prime}%
+Q_{42}^{\prime})/2,\nonumber \\
\tilde{Q}_{42}(q)  &  =(\mathrm{i}Q_{31}^{\prime}+\mathrm{i}Q_{41}^{\prime
}-\mathrm{i}Q_{32}^{\prime}-\mathrm{i}Q_{42}^{\prime})/2,\nonumber \\
\tilde{Q}_{43}(q)  &  =(-\mathrm{i}Q_{33}^{\prime}-\mathrm{i}Q_{43}^{\prime
}+\mathrm{i}Q_{34}^{\prime}+\mathrm{i}Q_{44}^{\prime})/2,\nonumber \\
\tilde{Q}_{44}(q)  &  =(Q_{33}^{\prime}+Q_{43}^{\prime}+Q_{34}^{\prime}%
+Q_{44}^{\prime})/2.
\end{align}

Next, we focus on the phase fluctuations. Upon integration of the amplitude
fields we obtain a phase-only effective action as
\begin{equation}
S_{1}=\frac{\beta}{2}\sum_{q}\theta^{\dag}(q)\tilde{Q}_{phase}(q)\theta(q),
\end{equation}
where $\theta^{\dag}(q)=(\theta_{A}(q),\theta_{B}(q))$. The matrix $\tilde
{Q}_{phase}$ corresponding to phase-phase fluctuation is given by
\begin{equation}
\tilde{Q}_{phase}(q)=\left(
\begin{array}
[c]{cc}%
\tilde{Q}_{ph11} & \tilde{Q}_{ph12}\\
\tilde{Q}_{ph21} & \tilde{Q}_{ph22}%
\end{array}
\right)
\end{equation}
where \begin{widetext}
\begin{align*}
\tilde{Q}_{ph11} &  =-(\tilde{Q}_{21}M_{11}^{\prime}\tilde{Q}_{12}+\tilde
{Q}_{24}M_{21}^{\prime}\tilde{Q}_{12}+\tilde{Q}_{21}M_{12}^{\prime}\tilde
{Q}_{42}+\tilde{Q}_{24}M_{22}^{\prime}\tilde{Q}_{42})+\tilde{Q}_{22}\\
\tilde{Q}_{ph12} &  =-(\tilde{Q}_{21}M_{11}^{\prime}\tilde{Q}_{13}+\tilde
{Q}_{24}M_{21}^{\prime}\tilde{Q}_{13}+\tilde{Q}_{21}M_{12}^{\prime}\tilde
{Q}_{43}+\tilde{Q}_{24}M_{22}^{\prime}\tilde{Q}_{43})+\tilde{Q}_{23}\\
\tilde{Q}_{ph21} &  =-(\tilde{Q}_{31}M_{11}^{\prime}\tilde{Q}_{12}+\tilde
{Q}_{34}M_{21}^{\prime}\tilde{Q}_{12}+\tilde{Q}_{31}M_{12}^{\prime}\tilde
{Q}_{42}+\tilde{Q}_{34}M_{22}^{\prime}\tilde{Q}_{42})+\tilde{Q}_{32}\\
\tilde{Q}_{ph22} &  =-(\tilde{Q}_{31}M_{11}^{\prime}\tilde{Q}_{13}+\tilde
{Q}_{34}M_{21}^{\prime}\tilde{Q}_{13}+\tilde{Q}_{31}M_{12}^{\prime}\tilde
{Q}_{43}+\tilde{Q}_{34}M_{22}^{\prime}\tilde{Q}_{43})+\tilde{Q}_{33}%
\end{align*}
\end{widetext}with the matrix $M^{\prime}(q)$
\begin{equation}
M^{\prime}(q)=\left(
\begin{array}
[c]{cc}%
\frac{\tilde{Q}_{44}(q)}{Z(q)} & \frac{-\tilde{Q}_{14}(q)}{Z(q)}\\
\frac{-\tilde{Q}_{41}(q)}{Z(q)} & \frac{\tilde{Q}_{11}(q)}{Z(q)}%
\end{array}
\right)
\end{equation}
where
\begin{equation}
Z(q)=\tilde{Q}_{11}(q)\tilde{Q}_{44}(q)-\tilde{Q}_{14}(q)\tilde{Q}_{41}(q).
\end{equation}
Finally, the total phase-only effective action becomes
\begin{equation}
S_{1}^{\prime}=\beta/2\sum_{q}\theta^{\dag}(q)\tilde{Q}_{phase}^{\prime
}(q)\theta(q)
\end{equation}
where the matrix $\tilde{Q}_{phase}^{\prime}(q)$ reads as
\begin{align}
\tilde{Q}_{phase}^{\prime}(q)  &  =\left(
\begin{array}
[c]{cc}%
\frac{1}{U}+\tilde{Q}_{ph11} & \tilde{Q}_{ph12}\\
\tilde{Q}_{ph21} & \frac{1}{U}+\tilde{Q}_{ph22}%
\end{array}
\right) \nonumber \\
&  =\left(
\begin{array}
[c]{cc}%
\tilde{Q}_{ph11}^{\prime} & \tilde{Q}_{ph12}^{\prime}\\
\tilde{Q}_{ph21}^{\prime} & \tilde{Q}_{ph22}^{\prime}%
\end{array}
\right)  .
\end{align}
The zero temperature superfluid density (phase stiffness) $\rho_{s}(0)$ is
obtained by examining the Goldstone mode action in the static limit. We
numerically extract the zero temperature superfluid stiffness $\rho_{s}(0)$ by
identifying
\begin{equation}
\sqrt{\tilde{Q}_{ph11}^{\prime}\tilde{Q}_{ph22}^{\prime}}-\sqrt{\tilde
{Q}_{ph12}^{\prime}\tilde{Q}_{ph21}^{\prime}}=[\sqrt{3}\rho_{s}%
(0)/2]\mathbf{q}^{2}%
\end{equation}
for $\left \vert \mathbf{q}\right \vert \rightarrow0$ \cite{zhao}.

Now we derive the superfluid density (the phase stiffness) for the TSC at zero
temperature $\rho_{s}(0)=Q.$ After obtaining the phase stiffness $\rho_{s}%
(0)$, the effective Lagrangian of the phase fluctuations is given by
\begin{equation}
L_{\mathrm{eff}}\simeq \frac{1}{2}\rho_{s}(0)(\mathbf{\bigtriangledown}%
\theta)^{2}\text{.}%
\end{equation}
where $\theta(x)$ denotes the phase fluctuations. See the results in Fig.15.

\begin{figure}[ptb]
\includegraphics[width=0.5\textwidth]{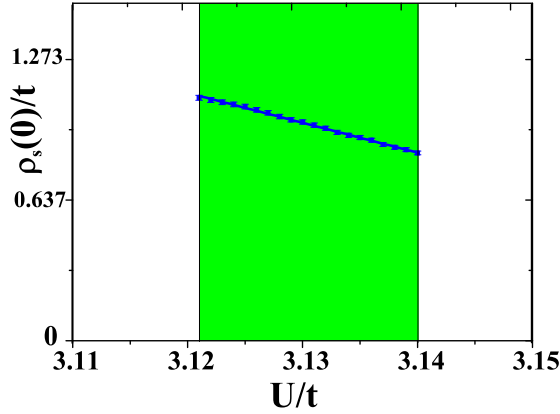}\caption{(Color online) The
phase stiffness of TSC order at $T=0$. The green region is the TSC order.}%
\label{kt}%
\end{figure}

\section{Physical realization}

In this end, we design an effective fermion model with $C=\pm2$ topological
superfluid (SF) as its ground state in an optical lattice. When two-component
fermions with repulsive interaction are put into a honeycomb optical lattice,
one can get an effective interacting Haldane model. It is easy to change the
potential barrier by varying the laser intensities to tune the Hamiltonian
parameters including the hopping strength ($t$-term), the staggered potential
($\varepsilon$-term) and the particle interaction ($U$-term).

We first design an optical lattice of the Haldane model. The Haldane model had
been proposed in the cold atoms with three blue detuned standing-wave lasers,
of which the optical potential is given by $V(x,y)=\sum_{j=1,2,3}V_{0}\sin
^{2}[k_{L}(x\cos \theta_{j}+y\sin \theta_{j})+\pi/2]$ where $V_{0}$ is the
potential amplitude, $\theta_{1}=\pi/3,$ $\theta_{2}=2\pi/3,$ $\theta_{3}=0$,
and $k_{L}$ is the optical wave vector in XY plane \cite{zhus}. When
two-component fermions are put into this honeycomb optical lattice, we may get
an effective Haldane model by applying the Raman laser beams
\begin{align}
\hat{H}_{\mathrm{H}}  &  =-t\sum \limits_{\left \langle {i,j}\right \rangle
,\sigma}\hat{c}_{i,\sigma}^{\dagger}\hat{c}_{j,\sigma}-t^{\prime}%
\sum \limits_{\left \langle \left \langle {i,j}\right \rangle \right \rangle
,\sigma}e^{\mathrm{i}\phi_{ij}}\hat{c}_{i,\sigma}^{\dagger}\hat{c}_{j,\sigma
}\nonumber \\
&  +\varepsilon \sum \limits_{i\in{A,}\sigma}\hat{c}_{i\sigma}^{\dagger}\hat
{c}_{i\sigma}-\varepsilon \sum \limits_{i\in{B,}\sigma}\hat{c}_{i\sigma
}^{\dagger}\hat{c}_{i\sigma}+h.c..
\end{align}
\ We introduce a complex phase $\phi_{ij}$ $\left(  \left \vert \phi
_{ij}\right \vert =\pi/2\right)  $ to the next nearest neighbor hopping, of
which the positive phase is set to be clockwise.\ To design a complex phase of
the next nearest neighbor hopping for two-component fermions generated by
gauge field on the optical lattice, Raman laser beams in XY plane are applied
with spacial-dependent Rabi frequencies as $\Omega_{0}\sin(\tilde{k}%
_{L}x+\frac{\pi}{4})e^{\mathrm{i}y\tilde{k}_{L}}$ and $\Omega_{0}\cos
(\tilde{k}_{L}x+\frac{\pi}{4})e^{-\mathrm{i}y\tilde{k}_{L}}$ ($\tilde{k}%
_{L}=2\pi/(3a)$) where $a$ denotes the length between nearest neighbor lattice
sites.\textbf{ }Then we get a laser-field-generated effective gauge field on
this honeycomb optical lattice as that given in Ref.\cite{shao}.

In addition, we consider a strong interaction via Feshbach resonance
technique$\hat{H}_{U}=-U\sum \limits_{i}\hat{n}_{i,\uparrow}\hat{n}%
_{i,\downarrow}$ where $U>0$ is the on-site attractive interaction strength
\cite{fesh,fesh1}. Thus, we may have an s-wave SF state by tuning the
interaction between fermions via Feshbach resonance technique.

Finally we get an interacting two-component fermions system in 2D honeycomb
optical lattice of the Haldane model as
\begin{equation}
\hat{H}=\hat{H}_{\mathrm{H}}+\hat{H}_{U}-\mu \sum \limits_{\left \langle
i,\sigma \right \rangle }\hat{c}_{i\sigma}^{\dagger}\hat{c}_{i\sigma}.
\label{haldane}%
\end{equation}
where $\mu$ denotes the chemical potential.

\section{Conclusion}

In this paper we studied the correlated Chern insulators by considering the
electrons on the Haldane's model with on-site negative-U interaction. We
obtained its properties by using the mean field theory and RPA\ approach. We
found that in the intermediate interaction region, the ground state becomes a
TSC with the Chern number $\pm2$. To characterize its topological properties,
we studied its edge states and the zero modes on the $\pi$-flux. In the end we
gave a proposal to realize such s-wave topological superconductor/superfluid
by putting two-component (two pseudo-spins) interacting fermions on a
honeycomb optical lattice.

\begin{acknowledgments}
This work is supported by the National Basic Research Program of China (973
Program) under grant No. 2012CB921704, 2011CB921803, 2011cba00102 and NFSC
Grant No. 11174035.
\end{acknowledgments}

\section{Appendix: The matrix $G_{0}$ of RPA approach to derive the phase
stiffness of TSC}

In this appendix, we give the details about the matrix $G_{0}$ ($=(G_{0}%
^{-1})^{-1}$) as
\begin{equation}
G_{0}=\left(
\begin{array}
[c]{cccc}%
G_{0,11} & G_{0,12} & G_{0,13} & G_{0,14}\\
G_{0,12} & G_{0,22} & G_{0,23} & G_{0,24}\\
G_{0,13}^{\ast} & G_{0,23}^{\ast} & G_{0,33} & G_{0,34}\\
G_{0,14}^{\ast} & G_{0,24}^{\ast} & G_{0,34} & G_{0,44}%
\end{array}
\right)  .
\end{equation}
Each element of $G_{0}$ can be divided into%
\begin{align}
G_{0,ij}  &  =\frac{A_{ij}}{\mathrm{i}\omega_{m}+E_{+}}+\frac{B_{ij}%
}{\mathrm{i}\omega_{m}-E_{+}}\nonumber \\
&  +\frac{C_{ij}}{\mathrm{i}\omega_{m}+E_{-}}+\frac{D_{ij}}{\mathrm{i}%
\omega_{m}-E_{-}}\nonumber \\
&  =\frac{\mathrm{numerator}}{\mathrm{denominator}}%
\end{align}
where the term "$\mathrm{numerator"}$ is\begin{widetext}
\begin{align*}
&  -\mathrm{i}\omega_{m}(A_{ij}+B_{ij}+C_{ij}+D_{ij})(\omega_{m}+\left \vert
\xi_{k}\right \vert ^{2}+\Delta_{0\text{ }}^{2}+\varepsilon^{2}+\gamma
_{\mathbf{k}}^{2})\\
&  +2\mathrm{i}\omega_{m}\gamma_{\mathbf{k}}(A_{ij}+B_{ij}-C_{ij}%
-D_{ij})\sqrt{\Delta_{0\text{ }}^{2}+\varepsilon^{2}}\\
&  +[(A_{ij}-B_{ij})E_{+}+(C_{ij}-D_{ij})E_{-}](\omega_{m}+\left \vert \xi
_{k}\right \vert ^{2}+\Delta_{0\text{ }}^{2}+\varepsilon^{2}+\gamma
_{\mathbf{k}}^{2})\\
&  +[-(A_{ij}-B_{ij})E_{+}+(C_{ij}-D_{ij})E_{-}]2\gamma_{\mathbf{k}}%
\sqrt{\Delta_{0\text{ }}^{2}+\varepsilon^{2}}%
\end{align*}
\end{widetext}and the term "$\mathrm{denominator"}$ reads
\[
(\mathrm{i}\omega_{m}+E_{+})(\mathrm{i}\omega_{m}-E_{+})(\mathrm{i}\omega
_{m}+E_{-})(\mathrm{i}\omega_{m}-E_{-}).
\]

For the element $G_{0,ij}$, we can express the coefficients as%
\begin{align}
A_{ij}  &  =\frac{M_{ij}+N_{ij}}{4}+\frac{P_{ij}-Q_{ij}}{4E_{+}},\nonumber \\
B_{ij}  &  =\frac{M_{ij}+N_{ij}}{4}-\frac{P_{ij}-Q_{ij}}{4E_{+}},\nonumber \\
C_{ij}  &  =\frac{M_{ij}-N_{ij}}{4}+\frac{P_{ij}+Q_{ij}}{4E_{-}},\nonumber \\
D_{ij}  &  =\frac{M_{ij}-N_{ij}}{4}-\frac{P_{ij}+Q_{ij}}{4E_{-}},
\end{align}
and the parameters for $M_{ij}$, $N_{ij}$, $P_{ij}$, $Q_{ij}$ are given as
follows: for $G_{0,11}$, we have
\begin{align}
M_{11}  &  =1,\nonumber \\
N_{11}  &  =\frac{\varepsilon}{\sqrt{\Delta_{0\text{ }}^{2}+\varepsilon^{2}}%
},\nonumber \\
P_{11}  &  =\gamma_{k}+\varepsilon,\nonumber \\
Q_{11}  &  =\frac{-(\varepsilon^{2}+\Delta_{0\text{ }}^{2}+\varepsilon
\gamma_{k})}{\sqrt{\Delta_{0\text{ }}^{2}+\varepsilon^{2}}};
\end{align}
for $G_{0,12}$, we have%
\begin{align}
M_{12}  &  =0,\nonumber \\
N_{12}  &  =\frac{-\Delta_{0\text{ }}}{\sqrt{\Delta_{0\text{ }}^{2}%
+\varepsilon^{2}}},\nonumber \\
P_{12}  &  =-\Delta_{0\text{ }},\nonumber \\
Q_{12}  &  =\frac{\Delta_{0\text{ }}\gamma_{k}}{\sqrt{\Delta_{0\text{ }}%
^{2}+\varepsilon^{2}}};
\end{align}
for $G_{0,13}$, we have%
\begin{align}
M_{13}  &  =0,\nonumber \\
N_{13}  &  =0,\nonumber \\
P_{13}  &  =-\xi_{k},\nonumber \\
Q_{13}  &  =\frac{\xi_{k}\varepsilon}{\sqrt{\Delta_{0\text{ }}^{2}%
+\varepsilon^{2}}};
\end{align}
for $G_{0,14}$, we have%
\begin{align}
M_{14}  &  =0,\nonumber \\
N_{14}  &  =0,\nonumber \\
P_{14}  &  =0,\nonumber \\
Q_{14}  &  =\frac{\xi_{k}\Delta_{0\text{ }}}{\sqrt{\Delta_{0\text{ }}%
^{2}+\varepsilon^{2}}};
\end{align}
for $G_{0,22}$, we have%
\begin{align}
M_{22}  &  =1,\nonumber \\
N_{22}  &  =\frac{-\varepsilon}{\sqrt{\Delta_{0\text{ }}^{2}+\varepsilon^{2}}%
},\nonumber \\
P_{22}  &  =\gamma_{k}-\varepsilon,\nonumber \\
Q_{22}  &  =\frac{-(\varepsilon^{2}+\Delta_{0\text{ }}^{2}-\varepsilon
\gamma_{k})}{\sqrt{\Delta_{0\text{ }}^{2}+\varepsilon^{2}}}.
\end{align}
for $G_{0,23}=-G_{0,14}$, we have%
\begin{align}
M_{23}  &  =0,\nonumber \\
N_{23}  &  =0,\nonumber \\
P_{23}  &  =0,\nonumber \\
Q_{23}  &  =\frac{-\xi_{k}\Delta_{0\text{ }}}{\sqrt{\Delta_{0\text{ }}%
^{2}+\varepsilon^{2}}};
\end{align}
for $G_{0,24}$, we have%
\begin{align}
M_{24}  &  =0,\nonumber \\
N_{24}  &  =0,\nonumber \\
P_{24}  &  =\xi_{k},\nonumber \\
Q_{24}  &  =\frac{\xi_{k}\varepsilon}{\sqrt{\Delta_{0\text{ }}^{2}%
+\varepsilon^{2}}};
\end{align}
for $G_{0,31}=G_{0,13}^{\ast}$%
\begin{align}
M_{31}  &  =0,\nonumber \\
N_{31}  &  =0,\nonumber \\
P_{31}  &  =-\xi_{\mathbf{k}}^{\ast},\nonumber \\
Q_{31}  &  =\frac{\xi_{k}^{\ast}\varepsilon}{\sqrt{\Delta_{0\text{ }}%
^{2}+\varepsilon^{2}}};
\end{align}
for $G_{0,32}=G_{0,23}^{\ast},$ we have%
\begin{align}
M_{32}  &  =0,\nonumber \\
N_{32}  &  =0,\nonumber \\
P_{32}  &  =0,\nonumber \\
Q_{32}  &  =\frac{-\xi_{k}^{\ast}\Delta_{0\text{ }}}{\sqrt{\Delta_{0\text{ }%
}^{2}+\varepsilon^{2}}},
\end{align}
for $G_{0,33}$, we have
\begin{align}
M_{33}  &  =1,\nonumber \\
N_{33}  &  =\frac{\varepsilon}{\sqrt{\Delta_{0\text{ }}^{2}+\varepsilon^{2}}%
},\nonumber \\
P_{33}  &  =-(\gamma_{k}+\varepsilon),\nonumber \\
Q_{33}  &  =\frac{\varepsilon^{2}+\Delta_{0\text{ }}^{2}+\varepsilon \gamma
_{k}}{\sqrt{\Delta_{0\text{ }}^{2}+\varepsilon^{2}}};
\end{align}
for $G_{0,34}$, we have%
\begin{align}
M_{34}  &  =0,\nonumber \\
N_{34}  &  =\frac{\Delta_{0\text{ }}}{\sqrt{\Delta_{0\text{ }}^{2}%
+\varepsilon^{2}}},\nonumber \\
P_{34}  &  =-\Delta_{0\text{ }},\nonumber \\
Q_{34}  &  =\frac{\Delta_{0\text{ }}\gamma_{k}}{\sqrt{\Delta_{0\text{ }}%
^{2}+\varepsilon^{2}}};
\end{align}
for $G_{0,41}=G_{0,14}^{\ast}$, we have%
\begin{align}
M_{41}  &  =0,\nonumber \\
N_{41}  &  =0,\nonumber \\
P_{41}  &  =0,\nonumber \\
Q_{41}  &  =\frac{\xi_{\mathbf{k}}^{\ast}\Delta_{0\text{ }}}{\sqrt
{\Delta_{0\text{ }}^{2}+\varepsilon^{2}}};
\end{align}
for $G_{0,42}=G_{0,24}^{\ast}$, we have
\begin{align}
M_{42}  &  =0,\nonumber \\
N_{42}  &  =0,\nonumber \\
P_{42}  &  =\xi_{k}^{\ast},\nonumber \\
Q_{42}  &  =\frac{\xi_{k}^{\ast}\varepsilon}{\sqrt{\Delta_{0\text{ }}%
^{2}+\varepsilon^{2}}};
\end{align}
for $G_{0,44}$, we have%
\begin{align}
M_{44}  &  =1,\nonumber \\
N_{44}  &  =\frac{-\varepsilon}{\sqrt{\Delta_{0\text{ }}^{2}+\varepsilon^{2}}%
},\nonumber \\
P_{44}  &  =-(\gamma_{k}-\varepsilon),\nonumber \\
Q_{44}  &  =\frac{\varepsilon^{2}+\Delta_{0\text{ }}^{2}-\varepsilon \gamma
_{k}}{\sqrt{\Delta_{0\text{ }}^{2}+\varepsilon^{2}}}.
\end{align}

\end{document}